\documentclass[12pt]{iopart}
\usepackage{iopams} \usepackage{nicefrac} \usepackage{graphicx}

\begin{document}

\date{\today}

\title[Anisotropic stars with quasi-local EoS]{Radial pulsations
and stability of anisotropic stars with quasi-local equation of state}

\author{Dubravko Horvat, Sa\v sa Iliji\'c and Anja Marunovi\'c}

\address{Department of Physics,
Faculty of Electrical Engineering and Computing,
University of Zagreb, Unska 3, HR-10\,000 Zagreb, Croatia}

\ead{sasa.ilijic@fer.hr}

\begin{abstract}
Quasi-local variables,
i.e.\ quantities whose values can be derived
from physics accessible within an arbitrarily small neighborhood
of a spacetime point,
are used to construct the equation of state (EoS)
for the anisotropic fluid in spherical symmetry.
One parameter families of equilibrium solutions are obtained
making it possible to assess stability properties
by means of the standard $M(R)$ method.
Normal modes of radial pulsation are computed as well
and are found to confirm the onset of instability
as predicted by the $M(R)$ method.
As an example, a stable configuration
with outwardly increasing energy density in the core is obtained
with a simple quasi-local extension of the polytropic EoS.
It is also found that the loss of stability occurs
at higher surface compactness
when the anisotropy of pressures is present.
\end{abstract}



\section{Introduction}

The most general spherically symmetric solution to Einstein equations
allows for the anisotropy of principal pressures;
the radial pressure may differ from the transverse pressure.
While the isotropic fluid with specified equation of state (EoS)
is the most common source for modeling Newtonian or relativistic stars,
anisotropy of principal pressures can be found
when the source is derived from field theories,
e.g.\ when scalar fields are considered as in boson stars \cite{SchuMi03},
or in configurations involving electrically charged matter.
One also finds that anisotropy of pressures is present
in the so called ``exotic solutions'' to Einstein equations
such as wormholes \cite{MoThoWH,VisserBook}
or gravastars \cite{VissGS2,DeBenedGS1}.
Interesting features of relativistic anisotropic spheres
were pointed out as early as 1933.\
by Lema{\^i}tre \cite{Lemaitre33,Lemaitre33reprint},
while ref.~\cite{BowLi74} is widely considered
as the beginning of the epoch of more active research,
see \cite{HerreraSantos97} for a review.
Recently, general algorithms for generating static anisotropic
solutions to Einstein equations have been formulated
(see \cite{Herrera07,KLake09} and references therein.)
However, whether such solutions are considered
more or less physically acceptable
depends on a number of criteria which include
fulfillment of various energy conditions of general relativity,
the requirement that the speed of sound in the fluid is subluminal,
and perhaps most importantly,
stability of the configurations with respect to perturbations.

The most notable technique for the analysis of stability
of isotropic relativistic stars with respect to radial perturbations
is that of Chandrasekhar \cite{Chandra64a}.
In this procedure the time dependent Einstein equations
are linearized around the equilibrium solution
to yield a linear wave equation for radial perturbations,
which together with appropriate boundary conditions
comprises an eigenvalue problem of the Sturm--Liouville type.
Positive eigenvalues are interpreted as
(squares of) frequencies of normal modes of radial pulsations,
while negative eigenvalues imply exponentially growing perturbations,
i.e.\ instability of the sphere
(for in-depth coverage see \cite{CatMeth,MTW73}).

Another field-proven technique, known as the $M(R)$ method,
can be applied to locate the onset of instability
in one-parameter families of equilibrium solutions to Einstein equations
satisfying the same EoS of cold matter \cite{CatMeth,HTWW65}.
Using the central energy density as the parameter
and increasing its value, the first maximum in the $M(R)$ curve
indicates the boundary between the stable and unstable configurations.
While more limited both in its scope
and insights into the dynamics of the system it can provide
relative to the full analysis of the normal modes,
the $M(R)$ technique is very simple in circumstances
where for the given EoS the equilibrium configurations
can be generated over a sufficiently wide parameter range.

In the context of anisotropic stars
the normal modes technique was applied to study the stability
of some specific configurations in \cite{HildeStei76}
and also \cite{DevGleis02,DevGleis03}.
However, the $M(R)$ method could not be straightforwardly applied
because most of the solution-generating techniques mentioned above
could not generate sequences of equilibrium configurations
corresponding to one fixed EoS.
The only framework outside of isotropic spheres
where the $M(R)$ method was successfully applied is that of elastic stars.
In the second paper in the series
beginning with \cite{Karlovini1,Karlovini2}
the $M(R)$ method was shown to predict instability
exactly at the configuration that has vanishing frequency
of the fundamental normal mode of radial pulsation,
i.e.\ the two methods were shown to give compatible results.

In this paper we will first show
how a sequence of anisotropic spheres of different masses and surface radii
can be generated from one given EoS.
This will enable us to use the $M(R)$ method to study stability,
and we will compare the results to those of the analysis of normal modes.
The paper begins (section~\ref{sec:eos})
with the discussion of the concept of the quasi-local EoS
for the anisotropic fluid in spherical symmetry.
In section~\ref{sec:waveq} we derive
the linear wave equation for radial perturbations
for a specific quasi-local EoS where the radial pressure
is described by the arbitrary barotropic EoS, $p=p(\rho)$,
while the anisotropy is taken to be the arbitrary function of
the energy density and the compactness, $\mu=2m/r$,
which is a quasi-local variable.
In section~\ref{sec:anisotropes}
we apply the above concepts to specific examples;
we study the structure and stability of solutions
obtained from the quasi-local EoS consisting of the
polytropic EoS governing the radial pressure
and the anisotropy of pressures taken to be bilinear
in the radial pressure and the compactness.
We close with a brief discussion of the results in section~\ref{sec:summary}.


\section{The equation of state \label{sec:eos}}

Energy--momentum tensor of the perfect fluid can be written as
$ \mathbf T = \rho \; \mathbf u \otimes \mathbf u
            + p \, ( \mathbf g + \mathbf u \otimes \mathbf u ) $,
where $\rho$ is the energy density
and $p$ is the isotropic pressure of the fluid,
$\mathbf u$ is its four-velocity normalized so that
$\mathbf u \cdot \mathbf u = -1$ and $\mathbf g$ is the metric tensor.
The combination $\mathbf g + \mathbf u \otimes \mathbf u$
is the projection tensor
onto the spatial hypersurface orthogonal to the fluid four-velocity.
The notion of the \emph{anisotropic} fluid
allows the pressures to differ among spatial directions.
In particular, in spherical symmetry,
the radial pressure which we denote by $p$
may differ from the transverse pressure which we denote by $q$.
To write down the energy--momentum tensor
of the anisotropic fluid in spherical symmetry,
in addition to the four-velocity $\mathbf u$,
one introduces the `radial' spatial unit vector $\mathbf k$
that is orthogonal to the four-velocity and has no angular components,
i.e.\ $ \mathbf k \cdot \mathbf k = 1 $, $\mathbf k \cdot \mathbf u = 0$.
The energy--momentum tensor of the anisotropic fluid
can now be written as
   \begin{equation} \label{eq:anisot}
   \mathbf T = \rho \; \mathbf u \otimes \mathbf u
             + p \; \mathbf k \otimes \mathbf k
             + q \, ( \mathbf g + \mathbf u \otimes \mathbf u
                                - \mathbf k \otimes \mathbf k )
   \end{equation}
where
$\mathbf g + \mathbf u \otimes \mathbf u - \mathbf k \otimes \mathbf k$
is the projection tensor onto the 2-surface
orthogonal to both $\mathbf u$ and $\mathbf k$.
Since the vector $\mathbf k$
is not defined at the centre of symmetry
the anisotropy of pressures must vanish at this point.

To illustrate the role of the anisotropy of pressures
in the dynamics of the fluid
it is worthwhile to consider the first law of thermodynamics:
   \begin{equation} \label{eq:firstlaw}
   \mathrm{d}(\rho V) = - \mathrm{d}W + T \, \mathrm{d}S,
   \end{equation}
where $V$ is the co-moving 3-volume of the fluid element,
$\rho$ and $T$ are respectively
the energy density and temperature of the fluid,
$S$ is the entropy and $W$ is the work done by the fluid element.
Assuming no energy flow
among the neighboring fluid elements we have $\mathrm{d}S=0$.
In the case of the isotropic pressure the work done by the fluid element
due to the change in its volume is $\mathrm{d}W = p\,\mathrm{d}V$.
Dividing by the element of the proper time
along the world line of the fluid element one can write
   \begin{equation} \label{eq:dWiso}
   \frac{\mathrm{d}W}{\mathrm{d}\tau}
   = p  \, \frac{ \mathrm{d}V }{ \mathrm{d}\tau }
   = pV \, (\boldsymbol \nabla \cdot \mathbf u)
   \end{equation}
where the relative change of the volume,
$(\mathrm{d}V/\mathrm{d}\tau)/V = {\boldsymbol \nabla} \cdot \mathbf u$,
can be recognized as the spacetime expansion scalar.
Applying the chain rule on the lhs\ of (\ref{eq:firstlaw}),
dividing by $\mathrm{d}\tau$ and using (\ref{eq:dWiso}),
for the isotropic fluid one obtains the well known relation
   \begin{equation} \label{eq:firstlawiso}
   {\mathbf u} \cdot {\boldsymbol \nabla} \rho =
   - ( \rho + p ) \, {\boldsymbol \nabla} \cdot \mathbf u.
   \end{equation}

Turning now to the work done by the element of the anisotropic fluid we must,
due to the difference among the principal pressures,
accordingly distinguish between different directions of expansion.
In spherical symmetry we distinguish between
the radial and the transverse expansion:
   \begin{equation}
   \boldsymbol \nabla \cdot \mathbf u =
   (\boldsymbol \nabla \cdot \mathbf u)_{\mathrm{rad.}}
   + (\boldsymbol \nabla \cdot \mathbf u)_{\mathrm{tr.}},
   \end{equation}
where the radial expansion can be written as
$ ({\boldsymbol \nabla} \cdot \mathbf u )_{\mathrm{rad.}} =
   \mathbf k \cdot ( \mathbf k \cdot \boldsymbol \nabla \mathbf u ) $.
The work due to the expansion of the element of the anisotropic fluid is
   \begin{equation}
   \mathrm{d}W
      = p \, \mathrm{d}V_{\mathrm{rad.}} + q \, \mathrm{d}V_{\mathrm{tr.}}
      = \big( pV \, (\boldsymbol \nabla \cdot \mathbf u)_{\mathrm{rad.}} +
      qV \, (\boldsymbol \nabla \cdot \mathbf u)_{\mathrm{tr.}} \big)
      \mathrm{d}\tau,
   \end{equation}
and the anisotropic analogue of (\ref{eq:firstlawiso}) follows as
   \begin{eqnarray} \label{eq:firstlawaniso}
   {\mathbf u} \cdot {\boldsymbol \nabla} \rho
   &= - \rho \, {\boldsymbol \nabla} \cdot \mathbf u
   - p \, ({\boldsymbol \nabla} \cdot \mathbf u)_{\mathrm{rad.}}
   - q \, ({\boldsymbol \nabla} \cdot \mathbf u)_{\mathrm{tr.}} \nonumber \\
   &= - ( \rho + q ) {\boldsymbol \nabla} \cdot \mathbf u
   - (p-q) ({\boldsymbol \nabla} \cdot \mathbf u)_{\mathrm{rad.}} \nonumber \\
   &= - ( \rho + q ) {\boldsymbol \nabla} \cdot \mathbf u
   - (p-q) \, \mathbf k \cdot ( \mathbf k \cdot \boldsymbol \nabla \mathbf u ).
   \end{eqnarray}
The last term on the rhs\ is proportional to the anisotropy
of the pressures and vanishes in case of the isotropic fluid,
thus reducing (\ref{eq:firstlawaniso}) to (\ref{eq:firstlawiso}).
However, (\ref{eq:firstlawaniso}) can also be obtained from the
energy--momentum tensor of the anisotropic fluid (\ref{eq:anisot})
by taking the component of the conservation law
$\boldsymbol \nabla \mathbf T = 0$
directed along the world line of the fluid element, 
$ \mathbf u \cdot \boldsymbol \nabla \mathbf T = 0$,
while the projection of $\boldsymbol \nabla \mathbf T = 0$
onto the hypersurface orthogonal to the four-velocity
gives the anisotropic version of the Euler equation
of relativistic hydrodynamics.

The equation of state (EoS) of the isotropic fluid is usually understood
as a relation among the energy density and the pressure of the fluid,
and possibly other \emph{local} fluid variables
such as temperature, baryon density, entropy per baryon, etc.
These variables are considered local
because their values refer to the state of the fluid
at the particular spacetime point,
rendering the EoS independent of any non-local information.
The local EoS describing the isotropic fluid
can be cast in the form of a single equation:
   \begin{equation}
   f(\rho,p,\dots)=0,
   \label{eq:isoeos}
   \end{equation}
dots representing additional local variables.
The simplest case is the barotropic EoS $f(\rho,p)=0$,
or as it is often written $p = p(\rho)$,
that can be taken to, e.g., describe the stellar matter
at the end of nuclear burning and at zero temperature.
It is well known \cite{Tolman39} that the barotropic EoS suffices to close
the static Einstein equations in spherical symmetry,
and as shown in \cite{RendSchmidt91},
a given barotropic EoS with pressure monotonically increasing in energy density
guarantees the existence of a family of solutions
with central energy density as the sole parameter.

In analogy to the case of the barotropic EoS for the isotropic fluid,
one could expect the simplest anisotropic EoS
to be of the form $p=p(\rho)$ for the radial,
and $q=q(\rho)$ for the transverse pressure.
However, a simple argument can be given to show
that such EoS is overly restrictive to allow
for a family of solutions with different central energy densities.
We begin by assuming the existence
of a static spherically symmetric solution
governed by the EoS of the form $p=p(\rho)$ and $q=q(\rho)$,
with energy density $\rho_0$ at $r=0$ (centre of symmetry)
and non-vanishing anisotropy of pressures at some $r=r_1>0$.
Since at $r=0$ the anisotropy of pressures must vanish
we have $p(\rho_0) = q(\rho_0)$.
At $r=r_1$ we have $p(\rho_1) \ne q(\rho_1)$ by assumption,
so it follows that $\rho_1 \ne \rho_0$.
If we now consider the possibility of constructing
the solution with central energy density equal to $\rho_1$,
we find that the condition of vanishing anisotropy of pressures
at $r=0$ cannot be satisfied.
The EoS for the anisotropic fluid must, therefore,
have more complex structure than what was just assumed.

The idea behind the \emph{quasi-local} EoS,
which we borrow from \cite{VissGS2,VissBH},
is to allow for the dependence on the quantities
that can be derived from the geometry at a given point of the spacetime.
In principle an arbitrarily small neighborhood of a spacetime point
suffices to measure the (orthonormal frame) components of the Riemann tensor.
These components, in turn, allow one to construct quantities
that can be considered as {quasi-local variables}.
In spherical symmetry the general quasi-local EoS
will have the form of a system
   \begin{equation} \label{eq:anisoeos}
   f_1(\rho,p,q,\dots;\mu,\dots)=0, \qquad
   f_2(\rho,p,q,\dots;\mu,\dots)=0,
   \end{equation}
where $\rho,p,q,\dots$ denote the local variables,
and $\mu,\dots$ denote the quasi-local variables.
The quasi-local variables of special interest
could be the curvature radius $r$ or the compactness $2m(r)/r$;
for orthonormal frame components
of the Riemann tensor in spherical symmetry see e.g.~\cite{VisserBook}.

In contrast to the local or quasi-local variables as defined above,
\emph{non-local} variables were defined
as arbitrary functionals over the whole, or some finite segment
of the $t=\mathit{const.}$ slice of the spacetime
\cite{HerNunPercoco99,HernandezNunez04}.
While in principle arbitrary functionals
can appear in the static limit of the EoS,
this concept might turn out to be problematic
when dynamics is considered.
Namely, it is not clear how a perturbation of the fluid
in the neighborhood of some spacetime point
would affect the value of an arbitrary functional,
and consequently the state of the fluid,
elsewhere in the spacetime.
The `immediate effect' of the perturbation on the functional
would violate the principle of causality,
while physically acceptable (causal) propagation
would require the input of additional physics,
implying that a non-local EoS alone does not fully describe the fluid.
However, the `average energy density up to $r$',
   \begin{equation}
   \overline \rho (r) = \frac{1}{\frac43 r^3 \pi}
   \int_0^r 4 r'^2 \pi \, \rho(r') \, \rmd r',
   \end{equation}
which is used as an example of a non-local variable
in \cite{HerNunPercoco99,HernandezNunez04} can,
in fact, be understood as a quasi-local variable
since $\overline \rho = 3 m / 4 r^3 \pi = 3 \mu / 8 \pi r^2 $,
where $\mu=2m/r$ and $r$ are acceptable quasi-local variables
\cite{VissGS2,VissBH}.


\section{The wave equation for radial perturbations \label{sec:waveq}}

Here we derive the linear wave equation
for the radial perturbations of the anisotropic fluid
governed by the specific form of the quasi-local EoS (\ref{eq:anisoeos}).
We assume that in the rest frame of the fluid its energy density $\rho$
and the pressures $p$ and $q$ are related by
   \begin{equation}
   p = p(\rho), \qquad q - p = a(\rho;\,\mu), 
   \label{eq:eos}
   \end{equation}
where the quasi-local variable $\mu$ is the compactness,
$\mu = 2m/r$, $m$ being the usual mass function.
The procedure that follows is a generalization
of the one originally carried out
for the isotropic case by Chandrasekhar \cite{Chandra64a},
while our notation is (to the extent possible)
compatible with Misner, Thorne and Wheeler \cite{MTW73}, Ch.~26.

The line element of the spherically symmetric spacetime
in the standard coordinates $t$ ,$r$, $\vartheta$ and $\varphi$ is
   \begin{equation}
   \rmd s^2 = - \rme^{2\Phi} \, \rmd t^2
             + \rme^{2\Lambda} \, \rmd r^2
             + r^2 \, \rmd \Omega^2 ,
   \label{eq:ds2}
   \end{equation}
where $\Phi=\Phi(t,r)$, $\Lambda=\Lambda(t,r)$,
and $\rmd \Omega^2 = \rmd\vartheta^2 + \sin^2\vartheta \, \rmd\varphi^2$
is the line element on the unit 2-sphere.
The metric function $\Lambda$ is related
to the compactness $\mu$ and to the mass function $m$ by
   \begin{equation} \label{eq:compactness}
   \mu = {2m}/r = 1 - \rme^{-2\Lambda} .
   \end{equation}

Denoting with $\xi=\xi(t,r)$ the radial displacement of a fluid element 
from its equilibrium position $r$ at time $t$,
the nonzero components of the four-velocity
normalized so that $u_{\mu}u^{\mu}=-1$ are
   \begin{equation}
   u^t = ( \rme^{2\Phi} - \rme^{2\Lambda} \dot\xi^2 )^{-1/2} , \qquad
   u^r = ( \rme^{2\Phi} - \rme^{2\Lambda} \dot\xi^2 )^{-1/2} \dot\xi ,
   \end{equation}
where $\dot\xi = \partial_t \xi = u^r / u^t$.
The nonzero components of the radial unit vector $k^{\mu}$
satisfying $k_{\mu}k^{\mu}=1$ and $k_{\mu}u^{\mu}=0$ are
   \begin{equation}
   k^t = \rme^{-\Phi+\Lambda}
         ( \rme^{2\Phi} - \rme^{2\Lambda} \dot\xi^2 )^{-1/2} \dot\xi, \qquad
   k^r = \rme^{\Phi-\Lambda}
         ( \rme^{2\Phi} - \rme^{2\Lambda} \dot\xi^2 )^{-1/2} .
   \end{equation}
According to (\ref{eq:anisot}) the energy--momentum tensor is
   \begin{equation}
   T^{\mu\nu} = (\rho+q) u^{\mu}u^{\nu} + q g^{\mu\nu} + (p-q) k^{\mu}k^{\nu} ,
   \label{eq:tmunu}
   \end{equation}
where $\rho=\rho(t,r)$ is the energy density,
$p=p(t,r)$ is the radial pressure
and $q=q(t,r)$ is the transverse pressure.

One now assumes the existence
of a static or equilibrium (time independent)
solution to the Einstein equations
and proceeds to study the dynamics of small perturbations.
At a fixed point in the coordinate grid
a quantity $f=f(t,r)$ is decomposed as
   \begin{equation}
   f(t,r) = f_0(r) + \delta f (t,r),
   \label{eq:euler}
   \end{equation}
where $f_0$ denotes the time independent equilibrium value of $f$
and $\delta f$ denotes the so called \emph{Eulerean} perturbation
(in equilibrium $\delta f=0$.)
In this way we decompose $\Phi$, $\Lambda$, $\rho$, $p$ and $q$,
and through the rest of this section we omit the explicit notation
of the dependence of variables on $t$ and $r$;
zero--subscripted equilibrium variables are understood to depend only on $r$,
while $\delta$--prefixed Eulerean perturbations depend on $t$ and $r$.

The Einstein equation, $ G^{\mu\nu} = 8\pi T^{\mu\nu} $,
for the metric (\ref{eq:ds2})
and the energy--momentum tensor (\ref{eq:tmunu})
can now be expanded in powers of the displacement $\xi$
and the Eulerean perturbations $\delta\Phi$, $\delta\Lambda$,
$\delta\rho$, $\delta p$ and $\delta q$.
The zero-order terms govern the equilibrium configurations:
   \numparts
   \begin{eqnarray}
   8 \pi \, \rho_0
      = r^{-2} ( - \rme^{-2\Lambda_0} ( 1 - 2r\Lambda_0') + 1 ) ,
      \label{eq:ee0tt} \\
   8 \pi \, p_0
      = r^{-2} ( \rme^{-2\Lambda_0} ( 1 + 2r\Phi_0'   ) - 1 ) ,
      \label{eq:ee0rr} \\
   8 \pi \, q_0
      = r^{-2} \rme^{-2\Lambda_0} (
         ( r \Phi_0' - r \Lambda_0' )( 1 + r \Phi_0' ) + r^2 \Phi_0'' ) ,
      \label{eq:ee0thth}
   \end{eqnarray}
   \endnumparts
while the first order terms
involve the fluid displacement and the the Eulerean perturbations:
   \numparts
   \begin{eqnarray}
   8\pi \, \delta\rho
      = 2 r^{-2} \rme^{-2\Lambda_0}
         ( ( 1 - 2r \Lambda_0' ) \delta\Lambda
            + r \, \delta\Lambda' ) ,
      \label{eq:ee1tt} \\
   8\pi \, \delta p
      = 2 r^{-2} \rme^{-2\Lambda_0}
         ( - ( 1 + 2r \Phi_0' ) \delta\Lambda + r \, \delta\Phi' ) ,
      \label{eq:ee1rr} \\
   8\pi \, \delta q
      = - \rme^{-2\Phi_0} \delta\ddot\Lambda + r^{-2} \rme^{-2\Lambda_0} \big(
            (1+2r\Phi_0'-r\Lambda_0')r \, \delta\Phi'
               + r^2 \delta\Phi'' \nonumber \\
                                \qquad \qquad \mbox{}
          - 2((r\Phi_0'-r\Lambda_0')(1+r\Phi_0')+r^2\Phi_0'')
          \delta\Lambda - (1+r\Phi_0')r \, \delta\Lambda' \big) ,
      \label{eq:ee1thth} \\
    (p_0 + \rho_0) \, \dot\xi
      = - 2r^{-2} \rme^{-2\Lambda_0} r \, \delta\dot\Lambda  
      \label{eq:ee1tr}
   \end{eqnarray}
   \endnumparts
(overdot denotes $\partial_t$, prime denotes $\partial_r$.)

From the EoS (\ref{eq:eos}) one may derive the relations
among the perturbation of the variables in the rest frame of the fluid.
These are called the \emph{Lagrangean} perturbations
and are denoted by the $\Delta$-prefix.
From (\ref{eq:eos}) we obtain
   \begin{equation}
   \Delta p = \frac{\rmd p}{\rmd \rho} \Delta \rho, \qquad
   \Delta q - \Delta p = \frac{\partial a}{\partial \rho} \Delta \rho
                       + \frac{\partial a}{\partial \mu} \Delta \mu .
   \end{equation}
A fluid rest frame (Lagrangean) perturbation of a quantity $\Delta f$ is,
for small perturbations,
related to the perturbation of the same quantity
seen in the coordinate frame (Eulerean perturbation) $\delta f$ with
   \begin{equation}
   \Delta f(t,r) =
      f(t,r+\xi(t,r)) - f_0(r) \simeq \delta f(r) + f_0'(r) \xi(t,r) .
   \end{equation}
Applying the above rule
we express the perturbations of the fluid variables
in terms of Eulerean perturbations, obtaining
   \begin{eqnarray}
\fl\delta p + p_0' \xi &=
      \left[\frac{\rmd p}{\rmd \rho}\right]_0
      (\delta \rho + \rho_0' \xi) , \label{eq:delp} \\
\fl\delta q + q_0' \xi &=
      \left( \left[\frac{\rmd p}{\rmd \rho}\right]_0
           + \left[\frac{\partial a}{\partial \rho}\right]_0 \right)
         (\delta \rho + \rho_0' \xi)
      + 2 \rme^{-2\Lambda_0}
         \left[ \frac{\partial a}{\partial \mu} \right]_0
         (\delta \Lambda + \Lambda_0' \xi) . \label{eq:delq}
   \end{eqnarray}
The quantities in brackets follow from the assumed EoS
and are to be evaluated at the equilibrium, as indicated
by the zero subscript.

The linearized Einstein equations
(\ref{eq:ee0tt}--\ref{eq:ee0thth}) and (\ref{eq:ee1tt}--\ref{eq:ee1tr}),
together with the perturbation
of the EoS (\ref{eq:delp}) and (\ref{eq:delq}),
constitute a system of equations that yields
the linear equation of motion for the fluid.
After a lengthy algebraic procedure aimed at
eliminating all Eulerean perturbations from the system
so that only the fluid displacement $\xi$ and its derivatives remain,
followed by the variable transformationr:
   \begin{equation} \label{eq:xi2zeta}
   \zeta(t,r) = r^2 \rme^{-\Phi_0(r)} \xi(t,r),
   \end{equation}
we arrive to the linear wave equation for the radial perturbations:
   \begin{equation} \label{eq:waveq}
      \ddot \zeta(t,r) =
      C_2(r) \, \zeta''(t,r) + C_1(r) \, \zeta'(t,r) + C_0(r) \, \zeta(t,r).
   \end{equation}
The coefficients $C_i(r)$ can be written as
   \numparts
   \begin{eqnarray}
 \fl  C_2 \, \rme^{ - 2 \Phi_0 + 2 \Lambda_0 } =
      & \left[ \frac{ \rmd p }{ \rmd \rho } \right]_0 , \label{eq:c2} \\[6pt]
 \fl  C_1 \, \rme^{ - 2 \Phi_0 + 2 \Lambda_0 } =
      & - \Phi_0'
        + \left( 2 \Phi_0' + \Lambda_0' - \frac2r \right)
          \left[ \frac{ \rmd p }{ \rmd \rho } \right]_0
        + \left[ \frac{ \rmd p }{ \rmd \rho } \right]_0' \nonumber \\
      & + \frac{ 4 (q_0-p_0) }{ r (\rho_0+p_0) }
          \left( 1 + \left[ \frac{\rmd p}{\rmd \rho} \right]_0 \right)
        - \frac{2}{r}
        \left[ \frac{ \partial a }{ \partial \rho } \right]_0 ,
                                                       \label{eq:c1} \\[6pt]
 \fl  C_0 \, \rme^{ - 2 \Phi_0 + 2 \Lambda_0 } =
      & \frac{ (1+r\Phi_0')^2 + \rme^{2\Lambda_0} - 2 }{ r^2 } \nonumber \\
      & + \frac{2(q_0-p_0)}{r^2(\rho_0+p_0)}
            \left( 2 r \Phi_0' + \big(3(r\Phi_0'-1)+r\Lambda_0'\big)
            \left[\frac{\rmd p}{\rmd\rho}\right]_0
        - 3 + r \left[\frac{\rmd p}{\rmd\rho}\right]_0' \right) \nonumber \\
      & - \frac{2 \Phi_0'}{r} \left[ \frac{\partial a}{\partial\rho}\right]_0
        + \frac{4 \rme^{-2\Lambda_0}}{r(\rho_0+p_0)}
          \left(
             \Lambda_0' \left[ \frac{\rmd p}{\rmd\rho} \right]_0 - \Phi_0'
          \right)
          \left[ \frac{\partial a}{\partial \mu} \right]_0.  \label{eq:c0}
   \end{eqnarray}
   \endnumparts
The quantities in brackets must be computed from the EoS
and evaluated using the equilibrium configuration variables.
Following the standard procedure
one assumes $\zeta(t,r) = u(r) \exp(\rmi \omega t)$
and the wave equation (\ref{eq:waveq}) reduces to
the ordinary differential equation
   \begin{equation} \label{eq:uode}
   - \omega^2 u = C_2 u'' + C_1 u' + C_0 u,
   \end{equation}
which together with the appropriate boundary conditions imposed on $u(r)$
(to be discussed below)
constitutes the boundary value problem with eigenvalue $\omega^2$.
(The function $u(r)$ should not be confused with the
four-velocity $u_{\mu}$.)
Solutions with $\omega^2>0$ are oscillatory in time
and therefore represent radial pulsations.
Solutions with $\omega^2<0$ may exponentially grow in time
and therefore imply instability of the equilibrium configuration
with respect to radial perturbations.

The boundary condition that we impose on $u(r)$ at $r=0$
follows from the observation that, due to symmetry,
the fluid element at $r=0$ cannot move. 
Therefore we have $\xi(t,r=0)=0$
which with (\ref{eq:xi2zeta}) gives $\zeta(t,r=0)=0$,
and finally $u(r=0) = 0$.
Additionally, the Lagrangean perturbation of the energy density
which can be written as
  \begin{equation}
  \Delta\rho = - \frac{\rme^{\Phi_0}}{r^2}
    \left(\frac2r(q_0-p_0) \zeta + (\rho_0+p_0) \zeta' \right)
  \end{equation}
must remain finite at $r=0$ from which it follows that
$u \propto r^3$ as $r \to 0$.
The situation is more delicate at the
surface of the star which we assume to be at finite radius $r=R$.
There we require that the fluid displacement $\xi(t,r=R)$
is finite at all times, which implies $u(r=R)$ finite.
Additionally,
we must ensure that at $r=R$ the Lagrangean perturbation
of the radial pressure, $\Delta p = [\rmd p / \rmd \rho]_0 \Delta\rho$,
vanishes at all times
as required by the junction conditions of the spherically symmetric
spacetime with the exterior Schwarzschild spacetime \cite{VisserBook}.
It is important to note that if $C_2\to 0$ as $r\to R$
then (\ref{eq:uode}) is singular at $r=R$.
Further analysis of the behaviour of $u$ as $r\to R$, therefore,
depends on the properties of the equilibrium configuration,
or more fundamentally, on the properties of the EoS
(see e.g.\ the discussion for the isotropic case in \cite{CatMeth}.)

By multiplying (\ref{eq:uode}) with the weight function
  \begin{equation} \label{eq:slw}
  W(r) = \exp \int^r
  \frac{C_1(\tilde r)-C_2'(\tilde r)}{C_2(\tilde r)}
  \, \rmd \tilde r
  \end{equation}
it assumes the Sturm--Liouville form, 
  \begin{equation}
  - \omega^2 W u = (P u')' + Q u ,
  \end{equation}
where $P=WC_2$ and $Q=WC_0$.
If $W>0$ over $[0,R]$ and $P$, $P'$, $Q$ and $W$ are continuous
we have a regular Sturm--Liouville problem
granting access to the main results of the Sturm--Liouville theory:
if the eigenvalues are ordered so that
$\omega_0^2 < \omega_1^2 < \dots < \omega_i^2 < \dots < \infty$,
then $\omega_i^2$ corresponds to the eigenfunction with $i$ nodes in $(0,R)$.
This implies that in order to test the stability of the configuration
with respect to radial perturbations
it is sufficient to compute the eigenvalue $\omega_0^2$
corresponding to the solution without nodes.
If $\omega_0^2>0$ then all solutions are oscillatory
and the whole equilibrium configuration is considered stable,
while if $\omega_0^2<0$ it is considered unstable.
(Solutions to the Sturm--Liouville eigenvalue problem
are usually referred to as `normal modes'
due to the orthogonality of the eigenfunctions
corresponding to different eigenvalues
with respect to the weight function $W(r)$.)

We end this section by noting that if one takes $a(\rho;\,\mu)=0$
all the relations derived above reduce to the well known relations
for the isotropic case (see e.g.\ \cite{CatMeth} or \cite{MTW73}, Ch.\ 26.)
We also note that instead of introducing
the adiabatic constant $\Gamma$ of the fluid
as it is done in the standard procedure in the isotropic case
where $\Gamma = (\rho+p)p^{-1}\,\rmd p/\rmd \rho$
we have, for clarity, preferred to work with the generic expressions
throughout the procedure.


\section{Example: anisotropic polytropes \label{sec:anisotropes}}

To give an example
where the concepts discussed so far are put to work
in this section we study the properties of spherically symmetric solutions
supported by the anisotropic fluid described with the quasi-local EoS:
  \begin{equation} \label{eq:eosfinal}
  p = p(\rho) = k \, \rho^{1+1/n}, \qquad 
  q - p = a ( \rho; \, \mu ) =  \alpha \, p(\rho) \, \mu .
  \end{equation}
The first relation is the polytropic EoS
that is assumed to relate the radial pressure to the energy density.
The parameter $n>0$ is known as the polytropic index.
We have chosen this EoS because the solutions where the polytropic EoS
governs the isotropic fluid, known as polytropes,
are probably the most thoroughly studied
self-gravitating objects in the literature,
both in Newtonian and in the relativistic regime \cite{Tooper65,HoredtBook}.
The second relation in (\ref{eq:eosfinal})
is the ansatz for the anisotropy of pressures
which we assume to be proportional
to the radial pressure and to the compactness.
The parameter $\alpha$ has the role of the anisotropy strength parameter.
An ansatz similar to this one has been used to generate the
gravastar solutions in \cite{DeBenedGS1}.
An important feature of this ansatz is that
the compactness which goes as $r^2$ when $r \to 0$
ensures the required vanishing of the anisotropy of pressures
at the centre of symmetry.
Further motivation for this ansatz was drawn from
astrophysical considerations that in the non-relativistic regime
where the compactness is much smaller than unity
the anisotropy of pressure is not expected to play an important role.
Also, when combined with the polytropic EoS assumed for the radial pressure,
this anisotropy ansatz ensures vanishing of the tangential pressure
at low energy densities, e.g.\ at or near the surface of the star.

The first part of our procedure is concerned
with the equilibrium configurations.
We begin by fixing the EoS
by choosing a triple of parameters
$k$, $n$ and $\alpha$ in (\ref{eq:eosfinal}),
and proceed to compute a sequence of static (equilibrium) solutions
to Einstein equations (\ref{eq:ee0tt}--\ref{eq:ee0thth})
corresponding to a sequence of central energy densities $\rho_0(0)$
from a certain range.
The equations to be solved for each $\rho_0(0)$
can be compactly arranged as the system of two coupled equations:
  \begin{eqnarray}
  m_0' & = 4 \pi \, r^2 \, \rho_0 \label{eq:mfn} , \\
  p_0' & = \frac2r (q_0 - p_0)
       - \frac{(\rho_0 + p_0)(m_0 + 4\pi \, r^3 p_0)}{r^2(1 - 2m_0/r)}
          \label{eq:atov},
  \end{eqnarray}
where (\ref{eq:mfn}) involves the mass function $m_0$ which is related
to the metric function $\Lambda_0$ and compactness $\mu_0$
by (\ref{eq:compactness}),
and (\ref{eq:atov}) is the anisotropic variant
of the well--known Tolman--Oppenheimer--Volkov (TOV) equation
(the zero-subscript indicates that the variables
correspond to the equilibrium configuration and prime denotes $\partial_r$.)
By using the EoS to express $p_0$ in terms of $\rho_0$,
and by writing $p_0'= [\rmd p / \rmd \rho]_0 \rho_0'$,
TOV is converted into the differential equation for
equilibrium energy density $\rho_0(r)$.
The solution to (\ref{eq:mfn}) and (\ref{eq:atov})
for the chosen EoS and central energy density $\rho_0(0)$
provides us with the surface radius $R$ and the total mass $M=m(R)$
of the equilibrium configuration.
Also, the $r$-dependence of $m_0$ and $\rho_0$ is known,
from which $p_0$, $q_0$, $\mu_0$ or $\Lambda_0$
can be readily obtained.
Finally, the metric function $\Phi_0$ is obtained by integrating
  \begin{equation}
  \Phi_0' = \frac{m_0 + 4\pi \, r^3 p_0}{r^2(1 - 2m_0/r)},
  \end{equation}
which completes the calculation of the equilibrium solution!

A sequence of equilibrium solutions corresponding to a given EoS,
i.e.\ fixed triple $n$, $k$, $\alpha$ in (\ref{eq:eosfinal}),
can be generated by varying the central energy density $\rho_0(0)$
as the only parameter.
Within the sequence the relation between the total mass $M$
and the surface radius $R$ can be studied.
It can be represented as a curve in the $M$ vs.\ $R$ plane
parametrized by the central energy density.
According to the $M(R)$ method \cite{HTWW65,CatMeth},
starting from the $\rho_0(0)\to0$ limit,
all solutions with central energy density
smaller than the one corresponding to the first maximum of the $M(R)$ curve
are stable with respect to radial perturbations.
In general, each extremum of the $M(R)$ curve
represents a critical configuration
in which some mode of radial pulsation
changes its stability properties.
Following the curve in the direction of increasing central energy density,
at a critical point (extremum) through which
the curve is making an anti-clockwise turn
a mode loses the stability;
in a clockwise turn through a critical point
a previously unstable mode becomes stable.

\begin{figure}
\begin{center}
\includegraphics{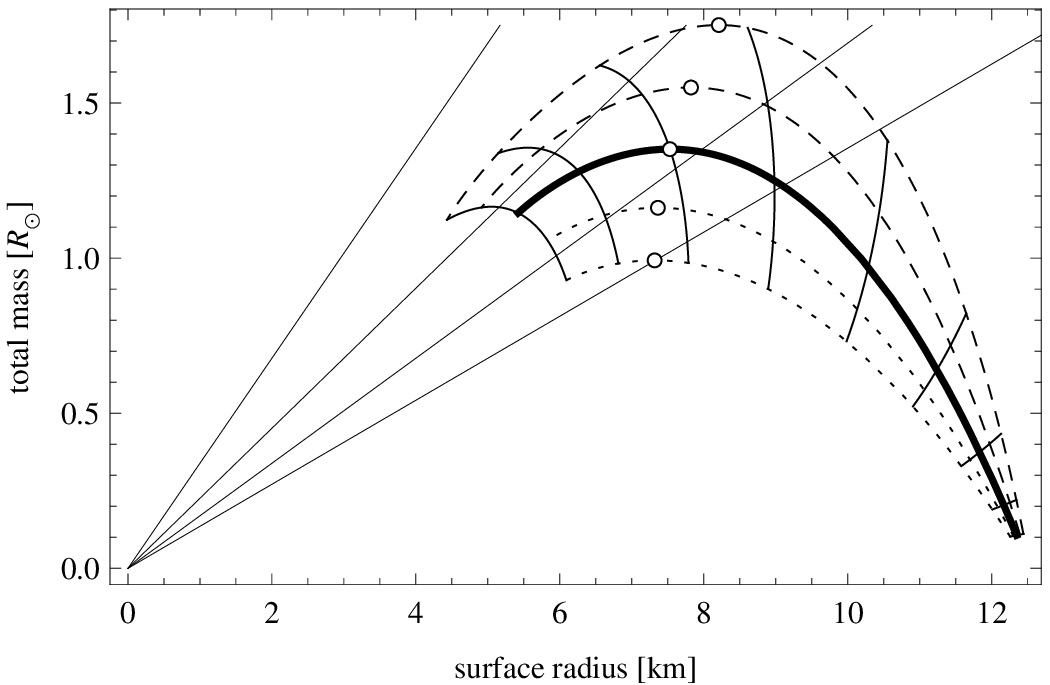}\\
\vskip 12pt \includegraphics{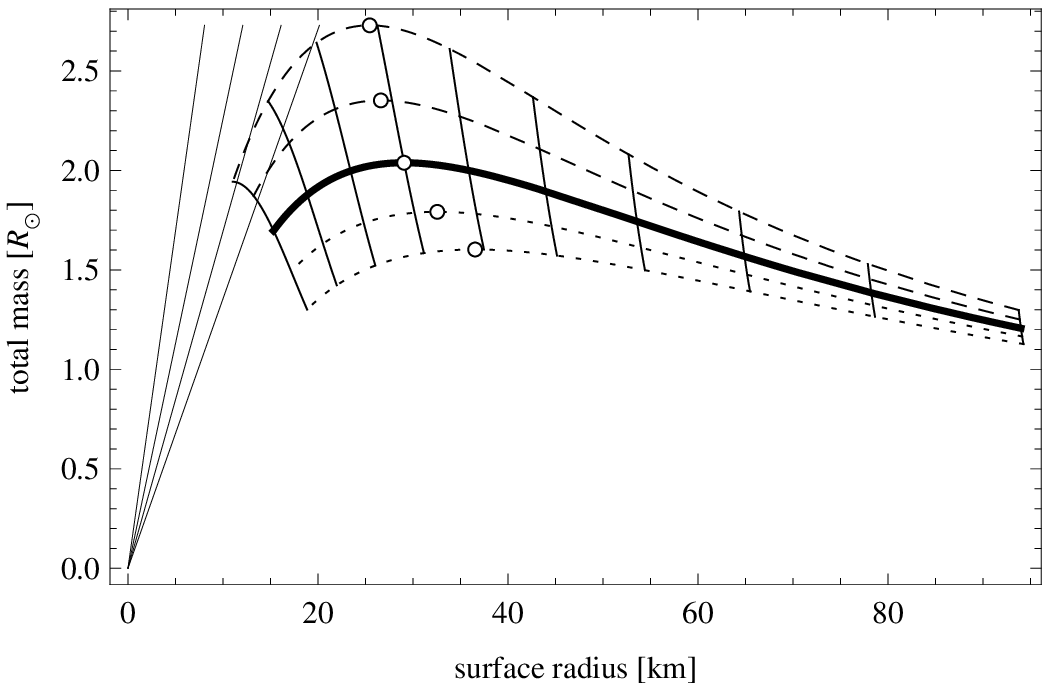}
\caption{\label{fig:one}
Mass--radius curves for anisotropic stars
with (radial) polytropic index $n=1$ and $k=100\,\mathrm{km}^{2}$ (upper plot),
and $n=2$, $k=5\,\mathrm{km}$ (lower plot):
the $M(R)$ curve for the $\alpha=0$ (isotropic) configuration
is shown with thick line,
for $\alpha=1,2$ with long-dashed,
and for $\alpha=-1,-2$ with short-dashed lines.
Solid curves connect the configurations sharing the same
central energy density and are drawn with step corresponding to factor 2
(central energy increases as one moves from right to left.)
Circles indicate the maxima of the $M(R)$ curves which coincide with
$\omega_0^2=0$ as obtained through the analysis of the normal modes.
Straight thin lines originating from $M=0=R$ indicate the locus of points
corresponding to surface compactness
$\mu_{\mathrm{s}}=2M/R=1,\nicefrac23,\nicefrac12,\nicefrac25$.
}
\end{center}
\end{figure}

Our first example uses the EoS
fixed by the triples $n=1$, $k=100\,\mathrm{km}^{2}$, $\alpha=0,\pm1,\pm2$.
This particular set of parameters
is chosen so that for the isotropic case ($\alpha=0$)
the $M(R)$ curve reproduces the reference configuration
of a relativistic polytrope presented in the table A.18
of Kokkotas and Ruoff \cite{KokkoRuoff}.
The $M(R)$ curve corresponding to $\alpha=0$
is shown with the thick line in the upper plot of figure~\ref{fig:one}.
The $M(R)$ curves corresponding to positive and negative
values of $\alpha$ show similar behaviour,
i.e.\ they have a maximum which we expect to
indicate the boundary between the stable and the unstable configurations.
The second example shown in the lower plot of figure~\ref{fig:one}
uses $n=2$, $k=5\,\mathrm{km}$ and $\alpha=0,\pm1,\pm2$.
In both examples the maxima of the $M(R)$ curves occur,
as compared to the isotropic configuration,
at higher surface compactness for $\alpha>0$.
This hints that anisotropy of pressures
with transverse pressure exceeding the radial pressure (our $\alpha>0$)
may be more efficient in supporting highly compact bodies
against gravitational collapse than in the opposite case (our $\alpha<0$.)
We can also observe that in our first example ($n=1$)
the maxima of the $M(R)$ curves occur at lower central energy densities
(as compared to the isotropic configurations) for positive $\alpha$,
while in the second example ($n=2$) this trend is reversed.

We now proceed to study the normal modes
of radial pulsations of the equilibrium configurations.
To this end we explicitly compute the solutions
to the boundary value problem consisting of (\ref{eq:uode})
with the boundary conditions $u(0)=0$ and $u(R)=\mathit{const.}$
This task is technically nontrivial
because with the EoS (\ref{eq:eosfinal})
the point $r=R$ is a regular singular point of (\ref{eq:uode}),
as can be shown by the following consideration.
The power expansion of the anisotropic TOV (\ref{eq:atov})
reveals that as $r\to R$,
  \begin{equation}
  \rho_0 \to
  \left( \frac{\mu_{\mathrm{s}}}{2k(n+1)(1-\mu_{\mathrm{s}})} \right)^n
  \left( 1 - \frac{r}{R} \right)^n,
  \end{equation}
where $\mu_{\mathrm{s}}=2M/R$ is the surface compactness.
With the above result
the asymptotic behaviour of the coefficients of
(\ref{eq:c2})-(\ref{eq:c0}) as $r\to R$ follows as
  \begin{equation}
\fl C_2\to \frac{(1-\mu_{\mathrm{s}})\mu_{\mathrm{s}}}{2n}
         \left(1-\frac{r}{R}\right), \quad
    C_1\to - \left( 1 + \frac1n \right)
         \frac{(1-\mu_{\mathrm{s}})\mu_{\mathrm{s}}}{2R}, \quad
    C_0\to \frac{(8-7\mu_{\mathrm{s}})\mu_{\mathrm{s}}}{4R^2} .
  \end{equation}
Due to the vanishing of $C_2$,
the point $r=R$ is a regular singular point of (\ref{eq:uode}),
which allows for unbounded solutions.
However, regularity of the solutions can be enforced
by integrating (\ref{eq:uode}) as an initial value problem
starting from $r=R$ with the initial condition $u(R)=\mathit{const.}$ and
  \begin{equation}
  u'(R) = - \frac{C_0 + \omega^2}{C_1} \, u(R)
        = \frac{8 - 7\mu_{\mathrm{s}} + (2R\omega)^2/\mu_{\mathrm{s}}}{
      2R(1+1/n)(1-\mu_{\mathrm{s}})} \, u(R) .
  \end{equation}
(We note that due to simplicity of the EoS (\ref{eq:eosfinal})
the above boundary condition coincides with equation (7c) of \cite{CatMeth}
that was derived assuming an asymptotically polytropic EoS
describing the fluid with isotropic pressures.)
Due to the robustness of the numerical method \cite{COLSYS},
finding the eigenvalue $\omega_i^2$ and the eigenfunction $u_i(r)$
satisfying the above boundary condition and also $u(0)=0$
with $i=0,1,\dots$ nodes in $(0,R)$
for any equilibrium configuration is now a simple task.
We are most interested in $i=0$ since,
as discussed in the preceding section,
if $\omega_0^2>0$ the equilibrium configuration is considered stable,
while if $\omega_0^2<0$ it is considered unstable.

For all solutions shown in figure~\ref{fig:one}
we computed the eigenvalue $\omega_0^2$ and observed
that as the central energy density increases ($n$, $k$ and $\alpha$ fixed)
it changes the sign from positive to negative exactly where,
within the remarkable numerical precision of the procedures,
the $M(R)$ curves exhibit their maxima.
In other words, the two methods we have applied
to assess the stability properties of the equilibrium configurations
obtained from the quasi-local EoS gave compatible results.

\begin{figure}
\begin{center}
\includegraphics{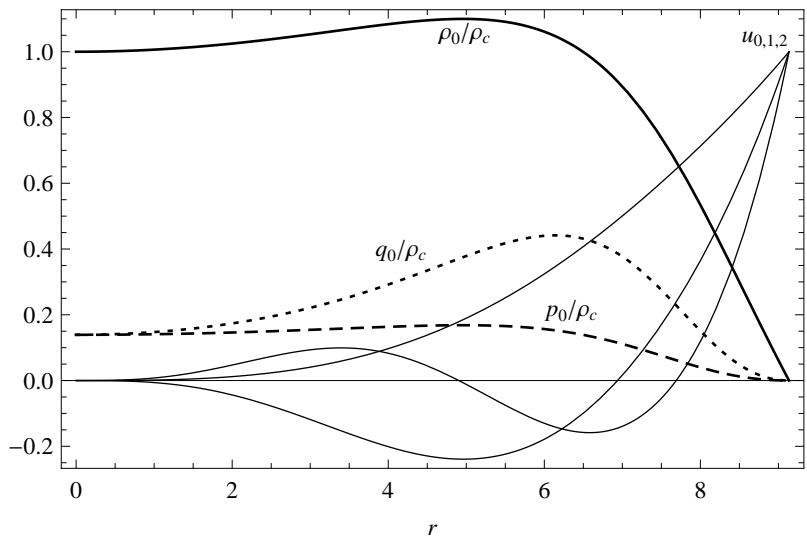}
\caption{\label{fig:two}
Critical (marginally un/stable) configuration with parameters
$k=100\,\mathrm{km}^{2}$, $n=1$, $\alpha=4$,
$\rho_0(0) = \rho_{\mathrm{c}}
\simeq 1.87\times10^{15}\,\mathrm{g}\,\mathrm{cm}^{-3}$,
$M\simeq2.14\,M_{\odot}$, $R\simeq9.14\,\mathrm{km}$, $2M/R\simeq0.69$.
Energy density (thick),
radial (long-dashed) and transverse pressures (short-dashed),
and the first three eigenfunctions (thin lines),
eigenfrequencies corresponding to the first three eigenfunctions are
$\nu_0\simeq0$, $\nu_1\simeq4.75\,\mathrm{kHz}$,
$\nu_2\simeq6.89\,\mathrm{kHz}$, respectively.}
\end{center}
\end{figure}

In figure~\ref{fig:two}
we show a highly compact configuration obtained with $n=1$,
$k=100\,\mathrm{km}^{2}$ and $\alpha=4$,
and central energy density for which $\omega_0^2=0$
(more details can be found in the figure caption.)
This configuration is therefore considered marginally stable.
An interesting feature that we can observe is the energy density
which is not maximal in the centre of the configuration,
but is outwardly increasing over the central region.

\section{Summary and outlook \label{sec:summary}}

We have shown that with the quasi-local variables
one can construct equations of state (EoS)
to describe the anisotropic fluid in spherical symmetry.
The quasi-local EoS may be used to generate
static anisotropic spheres over a wide range of radii and masses,
by varying the central energy density as the parameter.
We have also studied the stability of the configurations
with respect to radial perturbations.
The analysis of normal modes of radial pulsations 
and the $M(R)$ method gave identical results.
The numerical analysis of several examples revealed that
when the transverse pressure is greater than the radial pressure,
the maximal surface compactness ($2M/R$) of the spheres
that can be obtained before the onset of instability
is higher than that of isotropic spheres.
Another interesting feature is the outwardly increasing energy density
in the cores of some stable configurations,
despite of the fact that the radial pressure and the energy density
are related by the polytropic EoS (figure~\ref{fig:two}).
(Some may recall Weinberg's remark on isotropic spheres that
   ``it is difficult to imagine that a fluid sphere
     with a larger density near the surface
     then near the centre would be stable''
\cite{WeinbergBook};
we saw here that a similar structure
can be supported by anisotropic pressures,
whatsmore, it can be stable.)
Therefore, the concept of the quasi-local EoS
has been shown to be useful in constructing interesting models
of astrophysically plausible objects
such as highly compact stars,
especially since the $M(R)$ method
can be used to assess their stability properties.


\ack

We acknowledge the support
from the Croatian Ministry of Science
under the project 036-0982930-3144.


\section*{References}


\providecommand{\href}[2]{#2}\begingroup\raggedright\endgroup


\end{document}